# Electric Current Equilibrium in the Corona

Boris Filippov

*Pushkov Institute of Terrestrial Magnetism, Ionosphere and Radio Wave Propagation of the Russian Academy of Sciences (IZMIRAN), Troitsk, Moscow Region 142190, Russia*
*(e-mail:* bfilip@izmiran.ru*)*

**Abstract** A hyperbolic flux-tube configuration containing a null point below the flux rope is considered as a pre-eruptive state of coronal mass ejections that start simultaneously with flares. We demonstrate that this configuration is unstable and cannot exist for a long time in the solar corona. The inference follows from general equilibrium conditions and from analyzing simple models of the flux-rope equilibrium. A direct consequence of the stable flux-rope equilibrium in the corona are separatrices in the horizontal-field distribution in the chromosphere. They can be recognized as specific "herring-bone structures" in a chromospheric fibril pattern.

## 1. Introduction

Equilibrium conditions of magnetic-flux ropes have attracted the attention of solar physicists in connection with the problem of the onset and initiation of solar eruptive events, such as filament eruptions, flares, and coronal mass ejections (CMEs). When an electric current within a flux rope exceeds a certain threshold, a catastrophic loss of equilibrium follows (Van Tend and Kuperus, 1978; Hood and Priest, 1980; Molodenskii and Filippov, 1987; Martens and Kuin, 1989; Priest and Forbes, 1990; Forbes and Isenberg, 1991; Démoulin and Priest, 1988; Lin *et al*., 1998; Filippov, Gopalswamy, and Lozhechkin, 2001). This process was studied both analytically and numerically in 2D, 2.5D, and 3D geometry. The so-called torus instability can be generated by the hoop force (*e.g.* Bateman, 1978) in 3D geometry (Kliem and Török, 2006). Recently, Démoulin and Aulanier (2010) showed that the loss of equilibrium and the torus instability are two different views of the same physical mechanism.

Despite the progress in 3D numerical simulations, simple 2D models are still widely used in general to understand the fundamental properties of flux-rope equilibrium and stability. In qualitative, schematic considerations, different initial equilibrium configurations are presented. In particular, there are many cartoons (Pneuman, 1983; Malherbe and Priest, 1983; Anzer and Priest, 1985; Priest, 1990; Priest and Forbes, 1990) showing an inverse-polarity filament with an X-type singular point below it (Figure 1(b)). This figure-of-eght-type configuration is sometimes referred to as the Kuperus–Raadu model. However, in the Kuperus and Raadu original article (1974) only the configuration shown in Figure 1(a) was presented. This is also called the bald-patch separatrix surface configuration (BPS), while the former is called the hyperbolic flux-tube configuration (HFT: Titov, Priest, and Démoulin, 1993; Titov, Hornig, and Démoulin, 2002; Kliem, Török, and Forbes, 2011). The observation that the coronal mass ejection (CME) acceleration phase usually coincides with the soft X-ray flare rise phase, which was first demonstrated by Zhang *et al*. (2001), sometimes is interpreted as evidence of the HFT configuration in the corona prior to the eruption (Kliem, Török, and Forbes, 2011). Field-line reconnection, usually associated with a flare, can start at the X-type structure immediately after the flux-rope instability begins.

A null point can appear in the corona because of the complicated photospheric magnetic-field distribution (at least a quadrupolar structure). The two configurations shown in Figure 1 contain only a bipolar region and a coronal current. The null point exists in the corona because the fields of the bipolar arcade and that of the coronal current are superposed. In this article, the equilibrium conditions for a coronal electric current are considered. We conclude from the analysis that the configuration shown in Figure 1(b) is not in stable equilibrium and cannot be used as a pre-eruptive state for solar events.

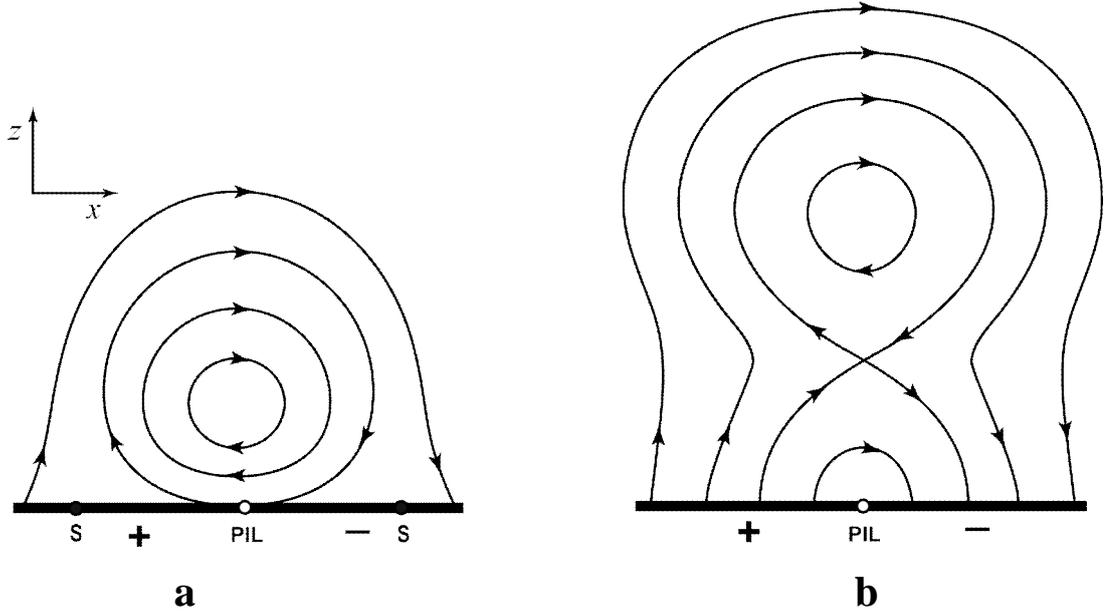

**Figure 1** Basic magnetic topology for the Van Tend and Kuperus (1978) model of prominence support (a) and the configuration with an X-point between a flux rope and the photosphere (b).

## 2. Integral Equilibrium Conditions for the Photospheric Field

In principle, equilibrium conditions in the corona are reflected in the distribution of the photospheric magnetic field. It is known that the MHD equation of momentum conservation can be written as (Landau, Lifshits, and Pitaevsky, 1984; Kuperus and Raadu, 1974)

$$\frac{\partial \rho v_i}{\partial t} = -\frac{\partial \Pi_{ik}}{\partial x_k}, \qquad (1)$$

where $x_i = (x, y, z)$ and summation over repeated indices is to be understood. $\Pi_{ik}$ is a symmetric tensor of second rank with the components

$$\Pi_{ik} = \rho\, v_i v_k + p\delta_{ik} - \frac{1}{4\pi}\left(B_i B_k - \frac{1}{2} B^2 \delta_{ik}\right). \qquad (2)$$

In equilibrium $\mathbf{v} = 0$, and in a low-$\beta$ coronal plasma we can neglect the gas-pressure term in Equation (2). Integrating Equation (1) with the zero left-hand side over the volume bounded by the surface $z = 0$ of the chromosphere (assumed here as a flat surface) and a hemisphere of radius $R$, we obtain after reducing the volume integral to a surface integral (Molodenskii and Filippov, 1989)

$$\oint \left(B_i B_k - \frac{1}{2} B^2 \delta_{ik}\right) ds_i = 0. \qquad (3)$$

Since for a concentrated source of the magnetic field, $B$ falls off at large distances as $R^{-3}$, we may integrate only over the surface $z = 0$ of the chromosphere. Correspondingly, the equilibrium conditions along the *x*-, *y*-, and *z*-axes for the chromospheric magnetic-field components are then given by

$$\int_{-\infty}^{\infty}\int_{-\infty}^{\infty} B_y B_z \, dxdy = 0, \qquad (4)$$

$$\int_{-\infty}^{\infty}\int_{-\infty}^{\infty} B_x B_z \, dxdy = 0, \qquad (5)$$

$$\int_{-\infty}^{\infty}\int_{-\infty}^{\infty} \left(B_z^2 - B_x^2 - B_y^2\right) dxdy = 0. \qquad (6)$$

These conditions for photospheric force-free fields were first obtained by Molodensky (1974). He also showed that these equations are satisfied in sunspots within the accuracy of measurements. Equation (6) describes the vertical equilibrium; this means that the mean-square value of the vertical field should be equal to the mean-square value of the horizontal field.

Let us represent field **B** as a sum of field **B**$_0$ of sub-photospheric sources and field **b** of coronal currents

$$\mathbf{B} = \mathbf{B}_0 + \mathbf{b}. \qquad (7)$$

Field **B**$_0$ is the potential field and satisfies Equations (4) – (6). Due to the photospheric diamagnetism

$$b_z(x,y,0) = 0. \qquad (8)$$

Substituting Equation (7) into (6) and using the boundary condition (8) and potentiality of **B**$_0$, we obtain (Molodenskii and Filippov, 1989)

$$\int_{-\infty}^{\infty}\int_{-\infty}^{\infty}\left[b_x(2B_{0x} + b_x) + b_y(2B_{0y} + b_y)\right] dxdy = 0. \qquad (9)$$

For the two-dimensional topology presented in Figure 1 we may set

$$b_y = 0 \qquad (10)$$

and to have the equilibrium condition in the form

$$\int_{-\infty}^{\infty} b_x(2B_{0x} + b_x) \, dx = 0, \qquad (11)$$

or

$$\int_{-\infty}^{\infty} b_x(B_{0x} + B_x) \, dx = 0. \qquad (12)$$

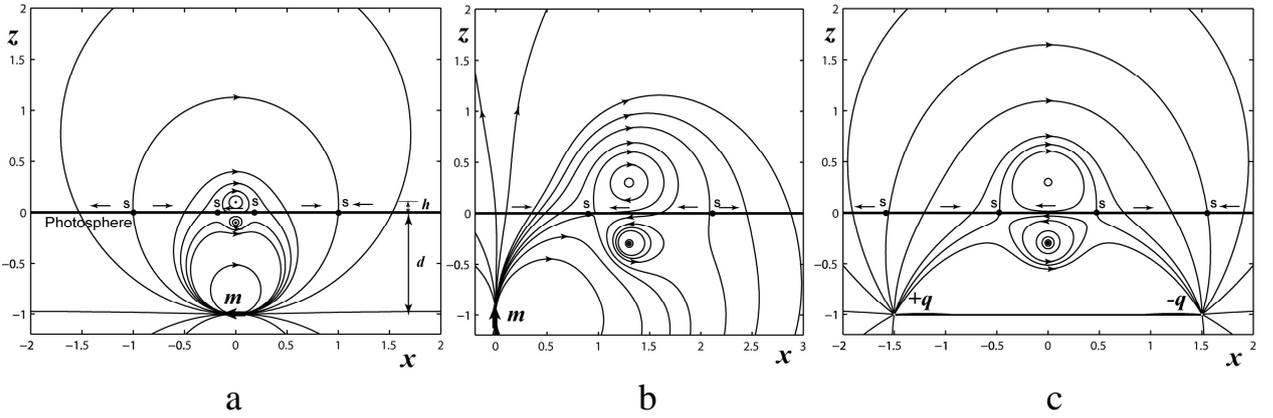

**Figure 2** Field lines of simple 2D models of line-current equilibrium in the field of a horizontal dipole (a), a vertical dipole (b), and two separated charges (c). Small arrows show the direction of the photospheric horizontal field. S denotes the separatrices.

For a single flux rope in the corona, as in Figure 1, the integration should be performed within limits on the order of the scale of the field [$b$], which is about the height [$h$] of the current above the chromosphere. The component $b_x$ has the same sign over the whole chromosphere. The component $B_{0x}$ can change sign, however, if the sub-photospheric currents are located rather deep below the surface at a depth $d$ (see Figure 2(a)), the lines of the $B_{0x}$ sign change are located at a distance on the order of $d$ away from a photospheric polarity-inversion line. When $h \ll d$, the sign of $B_{0x}$ is also constant within the integration region $-h < x < h$. Then, as seen from Equation (11), the signs of $B_{0x}$ and $b_x$ should be opposite, which is inherent for inverse-polarity filaments (Leroy, 1989; Paletou and Aulanier, 2003). Equation (12) needs the total field $B_x$ to change sign at least in some part of the integration region. This means the existence of separatrices S in the horizontal field distribution (Figure 1(a)). Just below the coronal current position, the direction of the entire field [$B_x$] is opposite to that of the sub-photospheric sources $B_{0x}$. Clearly, the configuration presented in Figure 1(b) does not meet this condition.

If $h \gg d$, the situation is not as clear. In this case, $B_{0x}$ changes sign within the integration region, therefore $B_x$, could in principle be unidirectional below the flux rope. However, in this condition, the flux-rope equilibrium cannot be stable (Molodenskii and Filippov, 1987; Priest and Forbes, 1990).

## 3. Distribution of the Photospheric Field in Simple Models

Condition (8) is satisfied if we introduce a virtual mirror electric current below the photosphere directed opposite to the coronal current (Kuperus and Raadu, 1974). The coronal current can find equilibrium only at an X-type null point (line), which occur when field $B_0$ of the sub-photospheric sources and field $b_m$ of the mirror current overlap (Figure 3). If $h \ll d$, the sign of the $B_x$ component below the flux rope down to the chromosphere is determined by the sign of the mirror current field $b_{mx}$, above the flux rope it is determined by the sign of $B_{0x}$. Obviously, only the left-hand configuration in Figure 1 meets this condition. The vertical equilibrium is stable because the Lorentz force acting on the flux rope is directed to the equilibrium position both below and above the null point.

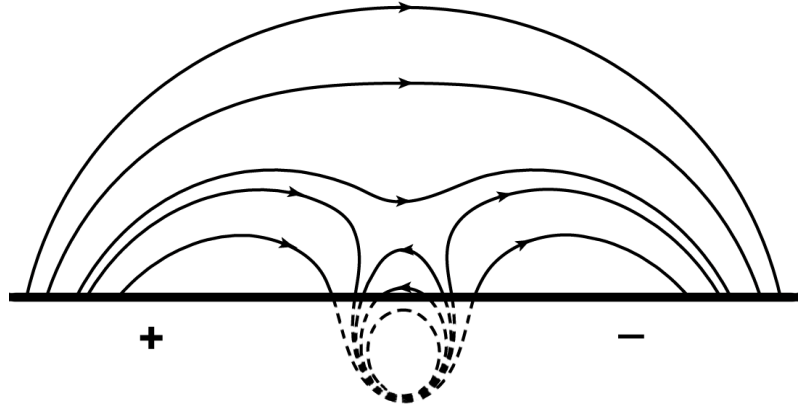

**Figure 3** Structure of the total field of the sub-photospheric sources and of the mirror current.

If $h \gg d$, the field near the null point changes direction. Below the null point, $B_0$ dominates, while above it $b_m$ determines the direction of the combined field. A possible equilibrium is now unstable, because the stronger field $B_0$ below the null point pulls the flux rope down, and the stronger field $b_m$ above the null point pushes it up.

*3.1. Horizontal Dipole*

Let us consider a simple 2D model describing the equilibrium of a line current in a dipole background magnetic field. It was developed by Molodenskii and Filippov (1987) for a vertical dipole and by Priest and Forbes (1990) for a horizontal dipole. The latter configuration is more symmetric and resembles Figure 1 more closely. If the line electric current [$I$] is directed along the horizontal axis $y$ and the $z$-axis is vertical, it generates a magnetic field in the upper hemisphere above a rigid conducting surface

$$b_x = \frac{2I}{c}\left(\frac{z-h}{x^2+(z-h)^2} - \frac{z+h}{x^2+(z+h)^2}\right), \tag{13}$$

$$b_z = \frac{2I\,x}{c}\left(\frac{1}{x^2+(z+h)^2} - \frac{1}{x^2+(z-h)^2}\right), \tag{14}$$

where $(0, h)$ are the current coordinates. The horizontal dipole $m$, located on the $y$-axis at the depth $d$ below the surface and directed opposite to the $x$-axis, creates the field

$$B_{0x} = -\frac{m\left(x^2-(z+d)^2\right)}{\left(x^2+(z+d)^2\right)^2}, \tag{15}$$

$$B_{0z} = -\frac{2mx(z+d)}{\left(x^2+(z+d)^2\right)^2}. \tag{16}$$

The equilibrium condition along the $z$-axis is

$$\frac{I^2}{c^2 h} - \frac{mI}{c(h+d)^2} = 0,$$

or

$$\frac{I}{mc} = \frac{h}{(h+d)^2}. \tag{17}$$

By substituting Equations (13) and (15) into the integrand of Equation (11), we find

$$\int_{-\infty}^{\infty} b_x^2 \, dx = \frac{16 I^2 h^2}{c^2} \int_{-\infty}^{\infty} \frac{dx}{(x^2+h^2)^2} = \frac{8\pi I^2}{hc^2}, \tag{18}$$

$$\int_{-\infty}^{\infty} 2 B_{0x} b_x \, dx = \frac{4 I h m}{c} \int_{-\infty}^{\infty} \frac{x^2 - d^2}{(x^2+h^2)(x^2+d^2)^2} \, dx. \tag{19}$$

The integration in the last expression can be easily performed if $h = d$,

$$\int_{-\infty}^{\infty} 2 B_{0x} b_x \, dx = \frac{8 I h m}{c} \int_{-\infty}^{\infty} \frac{x^2 - h^2}{(x^2+h^2)^3} \, dx = -\frac{2\pi I m}{h^3 c}. \tag{20}$$

Taking into account Equation (17), we see that the summation of Equations (18) and (20) yields zero, as the equilibrium condition (11) suggests.

Equilibrium is possible only if $I \leq mc/4d$. For every value of $I$ below the threshold, there are two equilibrium positions: the lower one, which is stable, and the upper state, which is unstable (Molodenskii and Filippov, 1987; Priest and Forbes, 1990). As the current $I$ increases, the lower equilibrium point grows, while the upper point shrinks. They merge at $h = d$, when $I = mc/4d$. A null point exists on the $z$-axis when $B_x(0, z_N) = 0$,

$$z_N = \frac{-4d + \sqrt{16d^2 - (4h^2 + (h+d)^2)(4d^2 - (h+d)^2)/h^2}}{(4h^2 + (h+d)^2)/h^2}. \tag{21}$$

There is no null point above the surface $z = 0$ (Figure 1(a)) at the lower equilibrium. A null point appears above the surface only when the height $h$ exceeds $d$, and it exists between the current and the surface at the upper equilibrium (Figure 1(b)). We recall that it is an unstable equilibrium. To summarize, a stable equilibrium is possible in this model only in the absence of null points in the corona.

The position of the separatrices S is determined by the equation $B_x(x, 0) = 0$, the solution of which is

$$x_s^2 = \frac{-8h^2 d^2 - (h+d)^2(h^2 - d^2) \pm \sqrt{[8h^2 d^2 + (h+d)^2(h^2 - d^2)]^2 - 4h^2 d^2 (4h^2 + (h+d)^2)(4d^2 - (h+d)^2)}}{2(4h^2 + (h+d)^2)}. \tag{22}$$

The value of $x_s$ is real only if $h \leq 0.23d$ or $h \geq d$. For low flux-rope heights there are four separatrices: two of them are located close to positions of the separatrices of the background field $B_0$: [$x_s = \pm d$]. The others are situated near the symmetry plane $x = 0$. If the height of the flux rope increases, the separatrices on each side of the symmetry plane move toward each other. They merge and disappear at $x_s = \pm 0.54d$ when $h = 0.23d$. For $0.23d < h < d$ there is no longer any separatrix in the $z = 0$ plane and everywhere $B_x < 0$, which corresponds to an "inverse" $B_x$ polarity below the flux rope. This means that the horizontal-field component is directed from negative to positive $B_z$ polarity. If $h$ increases above the altitude equal to $d$, a pair of separatrices diverging from the point $x = 0$ manifests the rising of a null point from the surface in the wake of the flux rope. In the limit $h \to \infty$, the separatrices approach the positions of the separatrices of the background field $B_0$, $x_s = \pm d$.

The separatrices are absent for the interval of heights $0.23d < h < d$ in this model because the dipole is located just below the polarity inversion line. This does not seem to reflect real solar conditions.

*3.2. Vertical Dipole*

For a vertical dipole, the flux rope may find equilibrium not on the vertical line above the dipole but on the line inclined at 45° to the vertical direction (Molodenskii and Filippov, 1987). The coordinates of the flux rope are now $x_0$ and $h$, and the horizontal components of fields are given by

$$b_x = \frac{2I}{c}\left(\frac{z-h}{(x-x_0)^2+(z-h)^2} - \frac{z+h}{(x-x_0)^2+(z+h)^2}\right), \quad (23)$$

$$B_{0x} = \frac{2mx(z+d)}{\left(x^2+(z+d)^2\right)^2}, \quad (24)$$

where $x_0 = h + d$. The strength of the equilibrium current for the given height becomes less by half, while the limiting height of the stable equilibrium is the same: $h = d$.

As for the horizontal dipole, a null point appears above the surface only if $h > d$. The position of separatrices is determined by the equation

$$h^2(x_s^2 + d^2)^2 - x_s d(h+d)^2\left[(x_s - h - d)^2 + h^2\right] = 0. \quad (25)$$

The numerical solution of the Equation (25) shows that the separatrix on the side of the dipole location is close to the position of the photospheric polarity-inversion line of the dipole magnetic field (Figure 2(b)). This separatrice exists for positions of the flux rope from close to the photospheric surface to heights much greater than $d$. The separatrice on the other side disappears when the height of the flux rope exceeds $0.45\ d$. For higher position of the flux rope, only one separatrix exists on the side of the stronger background magnetic field.

*3.3. Two Separated Charges*

Possibly more relevant to typical solar conditions is the background magnetic field represented by the two "charges" $\pm q$ (Figure 2(c))

$$B_{0x} = q\left[\frac{x+a}{(x+a)^2+(z+d)^2} - \frac{x-a}{(x-a)^2+(z+d)^2}\right], \quad (26)$$

where $\pm a$ are the $x$-coordinates of the charges.

The equilibrium of the flux rope requires the value of the electric current to be

$$I = \frac{2qahc}{a^2 + (h+d)^2}. \tag{27}$$

For every value of the electric current lower than the critical one

$$I_c = \frac{qac}{r+d}, \tag{28}$$

where $r = (a^2 + d^2)^{1/2}$; as in the dipole field, there are also two equilibrium positions

$$h_{1,2} = \frac{aqc}{I} - d \pm a\sqrt{\frac{q^2 c^2}{I^2} - \frac{2dqc}{I} - 1}. \tag{29}$$

The lower point is again stable, while the upper one is unstable. They merge at $h = r$, when $I = I_c$. The $z$-coordinate of a null point is given by

$$z_N = \frac{-8d + \sqrt{64d^2 - 4(5h^2 + 2hd + r^2)(3r^2 - 2hd - h^2)/h^2}}{2(5h^2 + 2hd + r^2)/h^2}. \tag{30}$$

The null point emerges from the surface after the flux rope reaches the height

$$h_N = \sqrt{3r^2 + d^2} - d = \sqrt{3a^2 + 4d^2} - d. \tag{31}$$

Evidently, it exceeds the limiting height of the stable equilibrium.

If $d \gg a$, the model tends to the case described in Subsection 3.1. In the reversed situation $d \ll a$, the limiting height of stability $h_c = a$ and $h_N = a\sqrt{3}$. In this approximation, there is a pair of separatrices above the charges $x_s = \pm a$ and a pair of $h$-dependent separatrices

$$x_s = \pm h\sqrt{\frac{3a^2 - h^2}{5h^2 + a^2}}. \tag{32}$$

The latter move away from the point $x = 0$ with the increase in flux-rope height until they reach the position $x_s = 0.6a$ when $h = 0.78a$. Then they return to the point $x = 0$, merge, and disappear when $h = a\sqrt{3}$. Obviously, this is the flux-rope height $h_N$ when the null point emerges from the surface. When the flux rope arrives at the point of the loss of stability $h = a$, the separatrices are still close to the greatest deviation from $x = 0$.

## 4. Chromospheric Fibril Pattern

Chromospheric fibrils show, with 180° ambiguity, the direction of the horizontal magnetic-field component in the chromosphere (Foukal, 1971; Zirin, 1972). Near filaments, they reflect the specific magnetic configuration that supports filament material high in the corona. A fibril pattern below a filament is so peculiar that it received a special name, "a filament channel" (Martres, Michard, and Soru-Escout, 1966; Gaizauskas *et al.*, 1997; Martin, 1998). Fibrils in the channel run nearly parallel to the polarity boundary, manifesting the presence of the strong-field component along the boundary. Filament barbs, which represent some prominent threads forming the filament's body, are parallel to the chromospheric fibrils just below the filament when viewed from above (Martin, 1998). Using information about the magnetic-polarity sense from a magnetogram and the field continuity, one can derive the direction of the horizontal magnetic field in a filtergram. Figure 4 clearly shows two lines that separate areas with opposite direction of a component perpendicular to the filament axis. Since a component parallel to the filament axis dominates in the filament channel, fibrils near the separatrices are aligned with the filament. They form a conspicuous "herring-bone structure" (Filippov, 1994, 1995). From this feature, separatrices near filaments can be easily recognized.

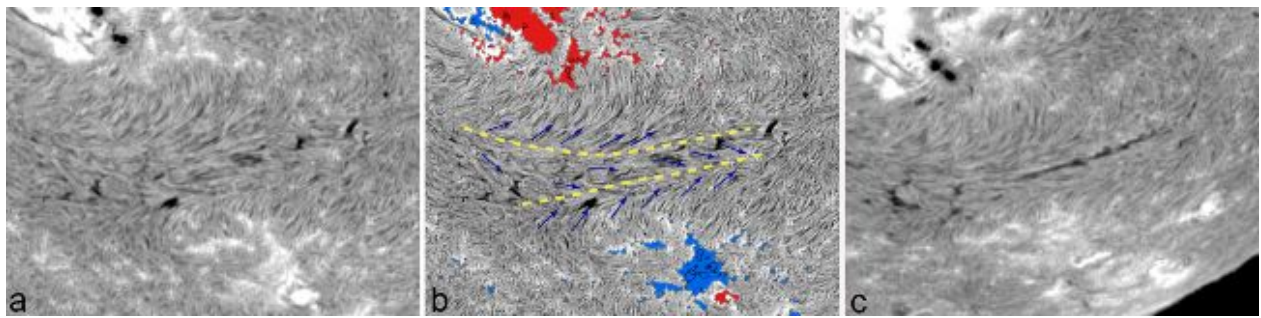

**Figure 4** (a) BBSO Hα filtergram on 10 November 2001. (b) The same filtergram with an unsharp–mask filter applied and with the photospheric magnetic-field concentrations stronger than ±50 gauss (G) from a SOHO/MDI magnetogram superposed. Dashed yellow lines represent separatrices, dark blue arrows show the orientation of the horizontal magnetic field component deduced from the fibril pattern. Red areas represent negative polarity, while blue areas represent positive polarity. (c) BBSO Hα filtergram on 12 November 2001. The filtergrams are rotated by 45º clockwise to arrange the filament axis to nearly horizontal in the frames. The images are 620" × 470" across (courtesy Big Bear Solar Observatory and SOHO/MDI consortium).

The filament in Figure 4(a) is very faint. Only a chain of small fragments is visible along the filament spine. However, owing to the filament transparency, the fibril pattern below it is more sharply defined. Nevertheless, the filament was visible every day during its passage across the disk because of the solar rotation. In the days after 10 November, the filament became more solid, although it was narrow and low.

The photospheric sources of the magnetic field in Figure 4 are concentrated and located rather far from the filament channel. The configuration is more similar to the model described in Section 3.3. The distance between the separatrices according to Equation (32) is about $2x_s = 2h\sqrt{3}$. In Figure 4(b), the distance between the separatrices is about 29 Mm. The height of the filament can be estimated in the filtergram taken on 14 November, when the filament was close to the limb. It is comparable with the width of the filament in Figure 4(c) and reaches nearly 9 Mm. Therefore, the foregoing relationship holds with good accuracy.

Separatrices near filaments can be recognized to a greater or lesser extent in many filtergrams (see Figure 5). Of course, the clearness of the structure depends on the regularity of the surrounding magnetic field. In some filtergrams in Figure 5, only one separatrix is clearly discernible. It could be related to a highly non-symmetric magnetic-field distribution relative to a polarity inversion line, a strong field on the one side and a faint field on the other side, as in the case of the vertical dipole in our models (Section 3.2), or a too strong a vertical-field component at the expected separatrix location, which prevents one from recognizing the horizontal field structure in a fibril pattern.

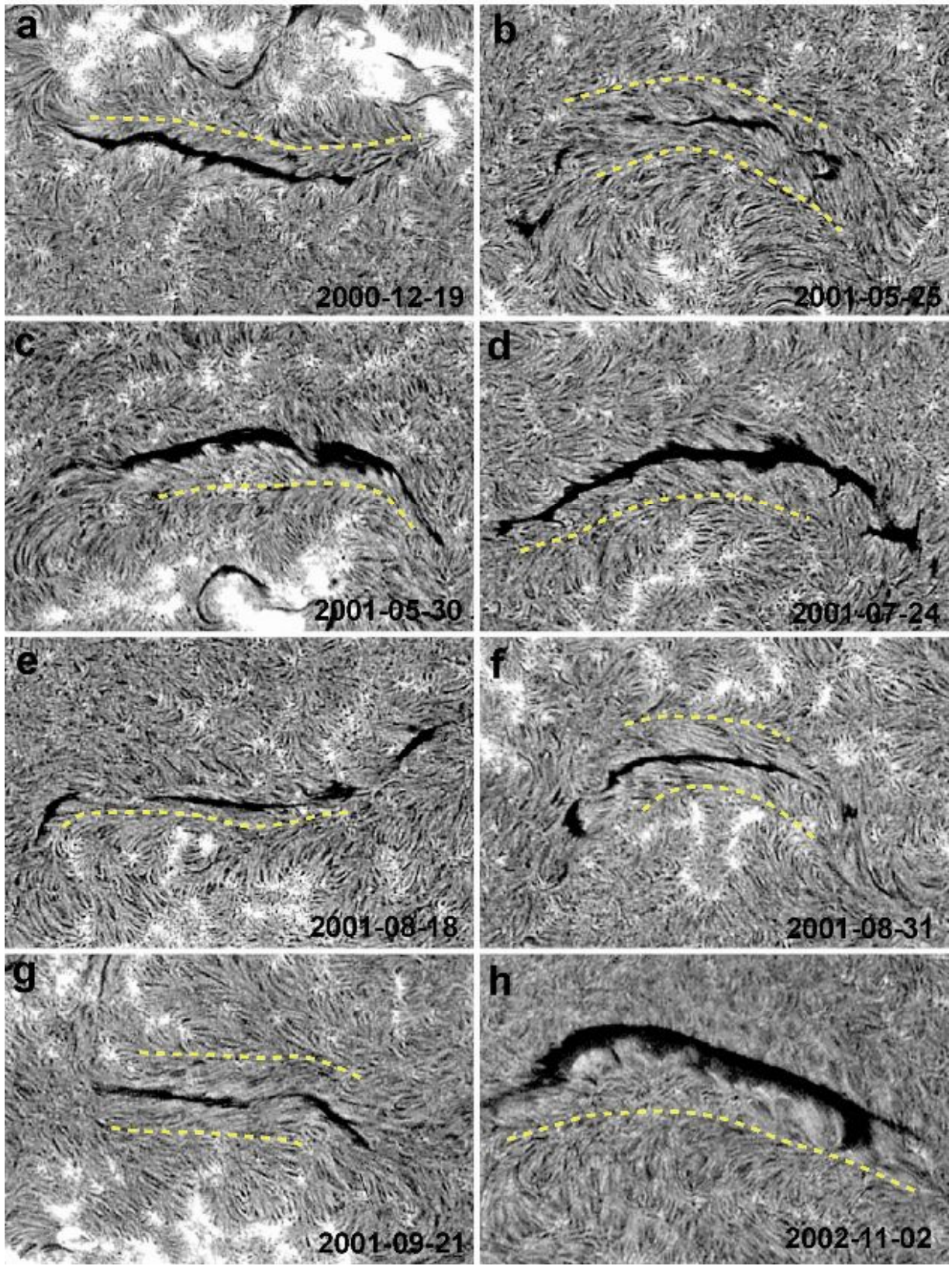

**Figure 5** BBSO Hα filtergrams showing separatrices in the fibril distribution near filaments. The size of each frame is 410" × 270". The filtergrams are rotated at different angles to arrange the filament axis to nearly horizontal in the frames. An unsharp mask filter was applied to highlight the structure (courtesy Big Bear Solar Observatory).

## 5. Summary and Conclusions

Magnetic configuration with a flux rope in the corona and a null point between the rope and the chromosphere (Figure 1(b)) is sometimes considered as a pre-eruptive state of CMEs. The advantage of this configuration is that field-line reconnection, usually associated with a flare, can start at the X-type structure immediately after the flux rope instability begins. However, we demonstrated that the figure-of-eight-type configuration is unstable and cannot exist for a long time in the solar corona. Therefore, it is not a suitable initial condition for CME formation. The configuration can appear during an eruption in a dynamic stage of the flux-rope evolution.

Equilibrium conditions in the corona are reflected in the distribution of the photospheric magnetic field. If the field is force-free, the mean-square value of the vertical field should be equal to the mean-square value of the horizontal field. But since only the line-of-sight component of magnetic fields in the photosphere is measured regularly, this relationship cannot be tested near filaments. One consequence from the general relationship is that the flux-rope field near the surface of the chromosphere must be opposite to the field of the sub-photospheric sources. For a wide variety of photospheric fields, the total horizontal field in the chromosphere changes direction at some lines, separatrices, parallel to PILs. We considered three simple models of flux-rope equilibrium and found that except in some special conditions the separatrices are present at the chromospheric level at both sides of a PIL.

The separatrices can be recognized in many filtergrams near filaments, to a greater or lesser extent. They could be considered as a necessary condition for stable flux-rope equilibrium, but lack of spatial resolution or the complicated photospheric-field structure do not allow one to find them near every filament. The bald-patch separatrix surface configuration also corresponds to recent vector magnetic-field measurements in a filament channel (López Ariste *et al.*, 2006).

**Acknowledgements** I thank the New Jersey Institute of Technology/Big Bear Solar Observatory and the Global High Resolution Hα Network for providing the data archive. Many thanks to the anonymous referee for constructive comments. This work was supported in part by the Russian Foundation for Basic Research (grants 12-02-00008 and 12-02-92692) and the Program # 22 of the Russian Academy of Sciences.